\documentclass[11pt]{article}

\def\be{\begin{equation}}
\def\ee{\end{equation}}
\def\te{\end{equation}}
\def\bea{\begin{eqnarray}}

\def\eea{\end{eqnarray}}
\def\tea{\end{eqnarray}}

\begin{document}

\title{Correlation Entropy of an Interacting Quantum Field and H-theorem for the
O(N) Model}
\author{E. A. Calzetta$\dagger$\thanks{%
Email: calzetta@df.uba.ar} and B. L. Hu$\ddagger$\thanks{%
Email: hub@physics.umd.edu} \\
$\dagger$ Departamento de Fisica, Facultad de Ciencias Exactas y Naturales\\
Universidad de Buenos Aires- Ciudad Universitaria,\\
1428 Buenos Aires, Argentina\\
$\ddagger$ Department of Physics, University of Maryland,\\
College Park, MD 20742, USA}
\date{(May 29, 2003, umdpp 03-050)}
\maketitle

\begin{abstract}
Following the paradigm of Boltzmann-BBGKY we propose a
correlation entropy (of the nth order) for an interacting quantum
field, obtained by `slaving' (truncation with causal
factorization) of the higher (n+1 th) order correlation functions
in the Schwinger-Dyson system of equations. This renders an
otherwise closed system effectively open where dissipation
arises. The concept of correlation entropy is useful for
addressing issues related to thermalization. As a small yet
important step in that direction we prove an H-theorem for the
correlation entropy of a quantum mechanical O(N) model with a
Closed Time Path Two Particle Irreducible Effective Action at the
level of Next-to-Leading-Order large N approximation. This model may be regarded as a field theory in $0$ space dimensions.
\end{abstract}

\section{Introduction}

The goal of this paper is two-fold: the proposal of a correlation
entropy for an interacting quantum field, and the proof of a
H-theorem for the quantum mechanical O(N) model
\cite{MACDH00,MDC01}, which may be regarded as a field theory in zero space dimensions. For the former, we follow the paradigm of
Boltzmann-BBGKY \cite{balescu} and propose a correlation entropy
(of the nth order) for an interacting quantum field \cite
{KB62,D84,CH88} obtained by `slaving' the higher (n+1 th) order
correlation functions in the Schwinger-Dyson system of equations
(See \cite{cddn,CH00} and below). We then derive the closed time
path (CTP) \cite{ctp} two particle irreducible (2PI) \cite{2pi}
effective action (EA) for this model up to the next-to-leading
order (NLO) in 1/N \cite{ON}. As a useful step towards addressing
issues related to the thermalization of interacting quantum
fields, we prove an H-theorem for the correlation entropy of a
quantum mechanical O(N) model at the CTP 2PI NLO level. We give an
introduction to the correlation hierarchy and the thermalization
issue in this section and a short description of different
definitions for the entropy of interacting quantum fields in the
next section, followed by a derivation of the H-theorem given in
the ensuing sections.\footnote{%
The following descriptive summary in this and the next section is
adapted from Chapter 5 of \cite {RamseyPhD}, where a preliminary
attempt to construct a correlation entropy was made but the proof
of the H-theorem for an interacting quantum field did not
materialize, and from \cite{CH02}.}

\subsection{Boltzmann-BBGKY Hierarchy and Schwinger-Dyson Equations}

We begin with nonequilibrium statistical mechanics. As is well
known, truncation of the BBGKY hierarchy at a finite nth order
yields a closed system \footnote{%
The truncation of the correlation hierarchy is not just an
arbitrary mathematical procedure, it reflects the fact that in
realistic conditions, measurement are of finite accuracy
associated with limited resolution of the instruments. Thus the
relevant physical degrees of freedom are often limited to the
lower end of the hierarchy, namely, the mean field and two-point
function.}. The system of equations describing the $n$-particle
distribution functions are time reversal invariant. When a causal
factorization condition such as Boltzmann's molecular chaos
assumption is imposed -- that is, assuming that the n+1 th
correlation function can be factorized into a product of n th
correlation functions initially but not finally (after the
collisions) -- time-irreversibility appears and an $H$-theorem
obtains. This type of \emph{coarse graining\/} of the hierarchy
i.e., truncation plus causal factorization into a set of coupled
equations for the $n$-particle distribution functions is called
`slaving' in the language of \cite {cddn,CH00}. Slaving of the
n+1 -particle distribution function renders an otherwise closed
system (of the n+1 th order correlation functions) an effectively
open system (of nth order correlation functions) and ushers in
the appearance of dissipative dynamics \cite{CH88}. Noise and
fluctuations \cite{HuPhysica,HuBanff} should also appear, as
required by the fluctuation-dissipation relation (FDR)
\cite{fdr}, now manifesting not just for an open system near
equilibrium (as depicted in linear response theory), but for an
effectively open system \cite{cddn,CH00}.

A familiar example is the Boltzmann equation: it is not enough to
truncate the BBGKY hierarchy to include only the one-particle and
two-particle distribution functions. One needs to make an
additional assumption that the two-particle distribution function
\emph{factorizes\/} into a product of single-particle
distribution functions before the collisions, i.e., invoking the
molecular chaos hypothesis. Of course the colliding partners
become correlated after interaction. This causal distinction is
the origin of the macroscopic arrow of time and the appearance of
dissipation in the Boltzmann paradigm. FDR calls for a rightful
place for fluctuations in the Boltzmann equation, as demonstrated
in the derivation of a stochastic Boltzmann equation
\cite{KacLog,Spohn}.

The situation in quantum kinetic field theory is completely analogous. By
quantum kinetic field theory (see, e.g., \cite{KB62,D84,CH88}), we are
referring to the hierarchy of coupled equations for the relativistic Wigner
function and its higher-correlation analogs, obtainable from the variation
of the master effective action \cite{cddn,CH00} whose variation yields the
Schwinger-Dyson equations. This is a quantum analogue of the BBGKY
hierarchy, expressed in a representation convenient for distinguishing
between microscopic (quantum field-theoretic) and macroscopic (transport and
relaxation) phenomena \footnote{%
It should be pointed out that in order to \emph{identify\/} the relativistic
Wigner function with a distribution function for quasiparticles, one must
show that the density matrix has \emph{decohered,\/} and this is neither
guaranteed nor required by the existence of a separation of macroscopic and
microscopic time scales \cite{Hab90}.}.

One may choose to work with a truncation of the hierarchy of the Wigner
function and its higher correlation analogs, or one may instead perform a
slaving of, for example, the Wigner-transformed four-point function, which
leads (within the context of perturbation theory) directly to the
relativistic Boltzmann equation \cite{CH88} and the usual $H$-theorem. The
truncation and subsequent slaving of the hierarchy within quantum kinetic
field theory can be carried out at any desired order, as dictated by the
initial conditions and relevant interactions.

\subsection{Thermalization in Interacting Quantum Fields}

The problem of thermalization in relativistic quantum fields has drawn much
attention over time, both in the attempt to understand the origin of
macroscopic irreversible behavior from microscopic theories, and for
practical considerations of nonequilibrium quantum field processes in the
early Universe and in relativistic heavy ion collisions \cite{S92}.

Let us begin with a discussion of the exact meaning of thermalization. In
the strictest sense an isolated system depicted by quantum field theory
undergoes unitary evolution and does not thermalize. However, one can still
ask meaningful questions such as whether certain correlation functions may
converge to their thermal forms in some well defined physical limit. We may
call this a weak thermalization condition.

In practical terms, even asking this type of questions may seem too
ambitious. As described above, the system of equations for the correlation
functions form an infinite hierarchy. When certain causal factorizable
boundary condition 
is mposed, the truncated subsystem will show signs of
irreversibility and a tendency to equilibrate. For interacting
fields we have to deal with a truncated Schwinger - Dyson
hierarchy to make the analysis possible. Moreover, nontrivial
(point) field theories are plagued by divergences which can only
be controlled by regularization and renormalization within some
perturbative scheme. Therefore the question of even weak
thermalization can make sense only within a chosen level of
approximation, such as the nPI of the correlation hierarchy, the
loop expansion, the $1/N$ expansion, or expansion in powers of a
coupling constant, etc.

There have been recent claims based on numerical evidence
\cite{BC01,B02} that at the next-to-leading order (NLO) Large $N$
approximation an interacting quantum field may show signs of
thermalization. In a recent report \cite{CH02} we consider a
$O(N)$ invariant scalar field of unbroken symmetry, develop the
CTP 2PI EA \cite{CH88} in powers of $1/N$, retaining up to the
next-to-leading order [$O(1)$] terms, and show that the only time
translation invariant solutions are thermal. Our analytic result
provides support for similar claims. Here, we present an
alternative approach by defining a correlation entropy and
showing the existence of an H-theorem in the simpler quantum
mechanical $O(N)$ model.

\section{Entropy of Interacting Quantum Fields}

Because entropy is of such fundamental importance, it is useful
to discuss how entropy can be defined for the various ways of
approximating the features and dynamics of a quantum field.
First, for a unitarily evolving quantum field theory whose
dynamics is a closed system (as opposed to an `effectively open
system') and
governed by the quantum Liouville equation, it is well known that the {%
von~Neumann entropy} of the density matrix,
\begin{equation}
S_{VN}=-Tr[\rho (t)\ln \rho (t)],
\end{equation}
is exactly conserved. If there is a justifiable separation of macroscopic
and microscopic time scales, one can adopt the theoretical framework of
quantum kinetic field theory. If one makes the assumption of factorization
(equivalently, slaving of the Wigner-transformed four-point function), one
obtains the relativistic Boltzmann equation in the binary collision
approximation. The Boltzmann entropy $S_{B}$ defined in terms of the phase
space distribution $f(k,X)$ for quasiparticles can in this case be shown to
satisfy a relativistic $H$-theorem \cite{degroot,CH88}. We want to
generalize this to a correlation entropy for interacting quantum fields.

However, in the case where there does \emph{not\/} exist such a
separation of time scales, how does one define the entropy of a
quantum field? For nonperturbative truncations of the dynamics of
interacting quantum fields, this is a nontrivial question
\cite{HuKan}. Intuitively, one expects that any coarse graining
which leads to an effectively open system with irreversible
dynamics will also lead to the growth of entropy. Formally these
operations can be systematically  expressed in terms of the
projection operator techniques \cite{ProjOp}. A projection
operator $P$ projects out the \emph{irrelevant\/} degrees of
freedom (thus going over to an open system) from the total system
described by the density operator $\rho$, yielding the reduced
density matrix $\rho _{R}$
\begin{equation}
\rho _{R}(t)=P\rho (t),
\end{equation}
There exists a well-developed formalism for deriving the equation of motion
of the \emph{relevant\/} degrees of freedom, and in terms of it, the
behavior of the coarse-grained entropy \cite{ProjOp,balescu}
\begin{equation}
S_{CG}=-Tr[\rho _{R}(t)\ln \rho _{R}(t)],  \label{eq-cge4}
\end{equation}
which will in general not be conserved. The projection operator formalism
can be used to express the slaving of higher correlation functions in the
correlation hierarchy. From it one can define an entropy in effectively open
systems (see, e.g., \cite{anastopoulos:1997a}). (So far it has only been
implemented within the framework of perturbation theory.) Another equally
powerful method adept to field theory is the Feynman-Vernon influence
functional formalism \cite{if,HuBanff} which has been used to treat entropy
in quantum open systems (see, e.g., \cite{KMH}).

\subsection{Entropy special to Choice of Basis or Representation}

We now consider the entropy functions for quantum fields,
beginning with the simpler yet more subtle case of a free field.
Historically this issue was, to our knowledge, first raised in the
context of entropy generation from particle creation for a free
quantum field in an expanding Universe \cite{Par69} due to the
parametric amplification of vacuum fluctuations. The focal point
is a wave equation with a time-dependent natural frequency for the
amplitude function of a normal mode. (The same condition arises
for an interacting field (such as the $\lambda \Phi ^{4}$ theory)
in the Hartree-Fock approximation or the O$(N)$ field theory at
leading order in the large-$N$ expansion \cite{ON}.) Since the
underlying dynamics is clearly unitary and time-reversal
invariant in this case, a suitable coarse graining leading to
entropy growth is not trivially evident. Hu and Pavon
\cite{HuPav} first made the observation that a coarse graining is
implicitly incorporated when one chooses to depict particle
numbers in the $n$-particle Fock (or ``N'') representation or to
depict phase coherence in the phase (or ``P'') representation.
Various proposals for coarse graining the dynamics of parametric
oscillators have followed
\cite{Kan88,EntProParCre,HKM,HKMP96,KME}. The language of
squeezed states is particularly useful for describing entropy
growth due to parametric particle creation \cite
{grischuk:1990a,matacz:1994a,HKM,KMH}. For our purposes, the
essential features of entropy growth due to parametric particle
creation which distinguish it from correlational entropy growth
(to be discussed below), are that parametric particle creation
depends sensitively on the choice of representation for the state
space of the parametric oscillators, and the specificity of the
initial conditions.

\subsection{Entropy from Projecting out Irrelevant Variables}

In contrast to entropy growth resulting from parametric
\textit{particle creation} from the vacuum, entropy growth due to
\textit{particle interactions} in quantum field theory
\cite{HuKan} has a very different physical origin. A coarse
graining scheme was proposed by Hu and Kandrup \cite{HuKan} for
these processes. Expressing an interacting quantum field in terms
of a collection of coupled parametric oscillators, their proposal
is to define a reduced density matrix by projecting the full
density operator onto each oscillator's single-oscillator Hilbert
space in turn,
\begin{equation}
{\gamma }(\vec{k})\equiv Tr_{\vec{k}^{\prime }\neq \vec{k}}\rho ,
\end{equation}
and defining the reduced density operator as the tensor product $\Pi $ of
the projected single-oscillator density operators ${\gamma }(\vec{k})$,
\begin{equation}
\rho _{R}\equiv \Pi _{\vec{k}}{\gamma }(\vec{k}).
\end{equation}
The coarse-grained (Hu-Kandrup) entropy is then just given by Eq.~(\ref
{eq-cge4}), from which we obtain
\begin{equation}
S_{HK}=-\sum_{\vec{k}}{Tr}[{\gamma }(\vec{k})\ln {\gamma
}(\vec{k})].
\end{equation}
It is interesting to observe that for a spatially
translation-invariant density matrix for a quantum field theory
which is Gaussian in the position basis, this entropy is just the
von~Neumann entropy of the full density matrix, because the
spatially translation-invariant Gaussian density matrix separates
into a product over density submatrices for each $\vec{k}$
oscillator. This projection (Hu-Kandrup) coarse graining, like the
correlation-hierarchy (Calzetta-Hu) coarse graining scheme, does
not choose or depend on a particular representation for the
single oscillator Hilbert space. It is sensitive to the
establishment of correlations through the explicit couplings.

\subsection{Entropy from Slaving the Higher Correlations: The CTP 2PI
effective action}

A general procedure has been presented for obtaining coupled
equations for correlation functions at any order $l$ in the
correlation hierarchy, which involves a truncation of the
\emph{master effective action\/} at a finite order in the loop
expansion \cite{cddn,CH00}. By working with an $l$ loop-order
truncation of the master effective action, one obtains a closed,
time-reversal invariant set of coupled equations for the first
$l+1$ correlation functions, $\hat{\phi},G,C_{3},\ldots
,C_{l+1}$. In general, the equation of motion for the highest
order correlation function will be linear, and thus can be
formally solved using Green's function methods. The existence of
a unique solution depends on supplying causal boundary
conditions. When the resulting solution for the highest
correlation function is  back-substituted into the evolution
equations for the other lower-order correlation functions, the
resulting dynamics is \emph{not\/} time-reversal invariant, and
generically dissipative. Thus, as was described before, with the
slaving of the higher-order (Wigner-transformed) correlation
function in quantum kinetic field theory, we have rendered a
closed system (the truncated equations for correlation functions)
into an \emph{effectively open system.\/} In addition to
dissipation, one expects that an effectively open system will
manifest noise/fluctuations, as shown in \cite{cddn,CH00} for the
case of the slaving of the four-point function to the two-point
function in the symmetry-unbroken $\lambda \Phi ^{4}$ field
theory. Thus a framework exists for exploring irreversibility and
fluctuations within the context of an unitary quantum field
theory, using the truncation and slaving of the correlation
hierarchy. The effectively open system framework is useful for
precisely those situations, where a separation of macroscopic and
microscopic time scales (which would permit an effective kinetic
theory description) does \emph{not\/} exist, such as is
encountered in the thermalization issue.

While it is certainly not the only coarse graining scheme which could be
applied to an interacting quantum field, the slaving of higher correlation
functions to lower-order correlation functions within a particular
truncation of the correlation hierarchy, as a particular coarse graining
method, has several important benefits. It can be implemented in a truly
nonperturbative fashion. This necessitates a nonperturbative resummation of
daisy graphs, which can be incorporated in the truncation/slaving of the
correlation hierarchy in a natural way.

In this study, we are interested in the growth of entropy due to the coarse
graining of the \emph{correlation hierarchy\/} by slaving a higher
correlation function. The simplest nonperturbative truncation of the
Schwinger-Dyson equations for the $\lambda \Phi^4$ field theory which
contains the time-dependent Hartree-Fock approximation
is the two-loop truncation of the master effective action, in
which only the mean field $\hat{\phi}$, the two-point function $G$, the
three-point function $C_3$ are dynamical. All higher order correlation
functions obey algebraic constraints, and can thus be expressed in terms of
the three dynamical correlation functions.

While this truncation of the Schwinger-Dyson equations is well-defined and
could in principle be solved, it is disadvantageous because, as stated
above, without some coarse graining, the system will not manifest
irreversibility and will not equilibrate. Therefore we slave the three-point
function to the mean field and two-point function, and thus arrive at an
effectively open system. In principle, a systematic analysis of the
coarse-grained dynamics of the mean field and two-point function should
include stochasticity \cite{cddn,CH00}, but noise is not our primary concern
here.

\section{The H theorem for the quantum mechanical $O\left( N\right) $ model}

\subsection{The Next-to-Leading order Large $N$ Approximation}

The number $N$ of replicas of essentially identical fields (like
the $N$ scalar fields in an $O(N)$ invariant theory, or the
$N^{2}-1$ gauge fields in a $SU(N)$ invariant non-abelian gauge
theory) suggests using $1/N$ as a natural small parameter, with a
well defined physical meaning. Unlike coupling constants, this is
not subjected to renormalization or radiative corrections. By
ordering the perturbative expansion in powers of this small
parameter, several nonperturbative effects (in terms of coupling
constants) may be systematically investigated.

The ability of the $1/N$ framework to address the nonperturbative
aspects of quantum field dynamics has motivated a detailed study
of the properties of these systems. In non-equilibrium
situations, this formalism has been applied to the dynamics of
symmetry breaking \cite {ON,CHKMPA94,CKMP95,BVHKP98,BVHS99} and
self-consistent semiclassical cosmological models
\cite{BVHS96,BCV01,RH97a,RH97b,RHS,GKLS97}.

In the case of the\ $O(N)$ invariant theory, in the presence of a nonzero
background field (or an external gravitational or electromagnetic field
interacting with the scalar field) we may distinguish the longitudinal
quantum \ fluctuations in the direction of the background field, in field
space, from the $N-1$ transverse (Goldstone or pion) fluctuations
perpendicular to it. To first order in $1/N$, the longitudinal fluctuations
drop out of the formalism, so we effectively are treating the background
field as classical. Likewise, quantum fluctuations of the external field are
overpowered by the fluctuations of the $N$ scalar field. In this way, the $%
1/N$ framework provides a systematics and a quantitative measure of the
semiclassical approximation \cite{HarHor81}.

To leading order ($LO$), the theory reduces to $N-1$ linear fields with a
time dependent mass, which depends on the background field and on the linear
fields themselves through a gap equation local in time. This depiction of
the dynamics agrees both with the Gaussian approximation for the density
matrix \cite{EJP88,MP89} and with the Hartree approximation \cite{HKMP96}.
The $LO$ $1/N$ theory is Hamiltonian \cite{HKMP96} and time- reversal
invariant. It simply does not thermalize. For example, if we set up
conditions where both the background field and the self-consistent mass are
space-time independent, then the particle numbers for each fluctuation mode
will be conserved. The existence of these conservation laws precludes
thermalization \cite{BW97}.

We note that the failure of the $LO$ approximation to describe
thermalization is indicative of a more general breakdown of the
approximation at later times, where effects of particle interaction
dominate. Both the distribution of energy among the field modes and the
phase relationships (or lack thereof) among them affect the way quantum
fluctuations react on the background or external fields. Therefore, from
physical considerations, one can say that a theory which does not describe
thermalization becomes unreliable for most other purposes as well \cite{S96}.

This is where the next to leading order ($NLO$) approximation enters. It has
been applied to quantum mechanics \cite{MACDH00,MDC01}, classical field
theory \cite{BW98,ABW00a,ABW00b,BCDM01} and quantum field theory in $1$
space dimension \cite{AB01}, being contrasted both to exact numerical
simulations of these systems, as well as against other approximations
purporting to go beyond $LO$. The $NLO$ has been shown to be an accurate
approximation, even at moderate values of $N.$

The $2PI$ formalism is also suitable for this question because, provided an
auxiliary field is cleverly introduced, the $2PI$ $CTP$ effective action can
be found in closed form at each order in $1/N$ \cite{ON,AABBS02}. We now
begin our study with a concrete model.

\subsection{O(N) $\lambda X^{4}$ Theory}

To fix the physical ideas, we shall discuss the problem in zero
space dimensions, namely, the  quantum mechanical O(N) model. The
system dynamics is described by the Hamiltonian with variables
$X_{A}$ and their conjugate momenta $P^{A}$, where $A,B$ are the
$O(N)$ group indices, with
\begin{equation}
H={\frac{1}{2}} \left\{ P^{B}P^{B}+M^{2}X_{B}X_{B}+\frac{\lambda }{4N}\left(
X_{B}X_{B}\right) ^{2}\right\}
\end{equation}
The classical action

\begin{equation}
S=\int dt\;{\frac{1}{2}}\left\{ \dot{X}_{B}\dot{X}_{B}-M^{2}X_{B}X_{B}-\frac{%
\lambda }{4N}\left( X_{B}X_{B}\right) ^{2}\right\}
\end{equation}
where $M_{0}^{2}$ and $\lambda _{0}$ are the mass parameter and
coupling constant. We rescale $X_{B}\equiv\sqrt{N} x_{B}$

\begin{equation}
S=N\int dt\;{\frac{1}{2}}\left\{ \dot{x}_{B}\dot{x}_{B}-M^{2}x_{B}x_{B}-%
\frac{\lambda }{4}\left( x_{B}x_{B}\right) ^{2}\right\}
\end{equation}
Discarding a constant term, we may rewrite the classical action as

\begin{equation}
S=N\int dt\;{\frac{1}{2}} \left\{ \dot{x}_{B}\dot{x}_{B}-\left[ \frac{M^{2}}{%
\sqrt{\lambda }}+\frac{\sqrt{\lambda }}{2}x_{B}x_{B}\right] ^{2}\right\}
\end{equation}

To set up the $1/N$ resummation scheme, it is customary to
introduce the auxiliary field $\chi $, writing

\begin{equation}
S=\frac{N}{2}\int \left\{ \dot{x}_{B}\dot{x}_{B}-\left[ \frac{M^{2}}{\sqrt{%
\lambda }}+\frac{\sqrt{\lambda }}{2}x_{B}x_{B}\right] ^{2}+\left[ \frac{%
M^{2}-\chi }{\sqrt{\lambda }}+\frac{\sqrt{\lambda }}{2}x_{B}x_{B}\right]
^{2}\right\}
\end{equation}
whence

\begin{equation}
S=N\int dt\;\left\{ {\frac{1}{2}}\dot{x}_{B}\dot{x}_{B}-\chi \left[ \frac{%
M^{2}}{\lambda }+\frac{x_{B}x_{B}}{2}\right] +\frac{1}{2\lambda }\chi
^{2}\right\} .  \label{classact}
\end{equation}
{}From now on, we consider $\chi $ and $x_{B}$ as fundamental
variables on equal footing.

Because of the $O(N)$ symmetry, the symmetric point must be a solution of
the equations of motion. For simplicity, we shall assume we are within this
symmetric phase, and treat $x_{B}$ as a quantum fluctuation. We also split
the auxiliary field $\chi =\bar{\chi}+\tilde{\chi}$ into a background field $%
\bar{\chi}$ and a fluctuation field $\tilde{\chi}$. The action becomes

\begin{equation}
S=S_{0}+S_{1}+S_{2}+S_{3}
\end{equation}
$S_{0}$ is just the classical action evaluated at $x_{B}=0,$ $\chi =\bar{\chi%
}$:

\begin{equation}
S_{0}=\frac{N}{\lambda }\int dt\;\left\{ \frac{1}{2}\bar{\chi}^{2}-M^{2}\bar{%
\chi}\right\}
\end{equation}
$S_{1}$ contains terms linear in $\tilde{\chi}$ and can be set to
zero by a suitable choice of the background field $\bar{\chi}$:

\begin{equation}
S_{1}=\frac{N}{\lambda }\int dt\;\left\{ \bar{\chi}-M^{2}\right\} \tilde{\chi%
}
\end{equation}
$S_{2}$ contains the quadratic terms and yields the tree - level
inverse propagators,

\begin{equation}
S_{2}= N \int dt\;\left\{ {\frac{1}{2}}\dot{x}_{B}\dot{x}_{B}-\frac{\bar{\chi%
}}{2}x_{B}x_{B}+\frac{1}{2\lambda }\tilde{\chi}^{2}\right\}
\end{equation}
Finally $S_{3}$ contains the bare vertex

\begin{equation}
S_{3}=\left( \frac{-N}{2}\right) \int d^{d}x\;\left\{ \tilde{\chi}%
x_{B}x_{B}\right\}.
\end{equation}

To write the $2PI$ $CTP$ $EA$ we double the degrees of freedom,
incorporating a branch label $a=1,2$ (for simplicity, if not
explicitly written, we assume that the label $a$ also contains
the time branch, i.e.,  $x^{Aa}\equiv x^{Aa}\left( t_{a}\right)
$). We also introduce propagators $G^{Aa,Bb}$ for the path ordered
expectation values

\begin{equation}
G^{Aa,Bb}=\left\langle x^{Aa}x^{Bb}\right\rangle  \label{eq12}
\end{equation}
and $F^{ab}$ for

\begin{equation}
F^{ab}=\left\langle \tilde{\chi}^{a}\tilde{\chi}^{b}\right\rangle
\end{equation}
Because of symmetry, it is not necessary to introduce a mixed propagator, for $%
\left\langle \tilde{\chi}^{a}x^{Bb}\right\rangle \equiv 0$. The $2PI$ $CTP$ $%
EA$ reads

\begin{eqnarray}
\Gamma &=&S_{0}\left[ \bar{\chi}^{1}\right] -S_{0}\left[ \bar{\chi}%
^{2}\right]  \nonumber \\
&&+\frac{1}{2}\int dudv\;\left\{ D_{Aa,Bb}(u,v)G^{Aa,Bb}(u,v)+\frac{N}{%
\lambda _{0}}c_{ab}\delta (u,v)F^{ab}\left( u,v\right) \right\}  \nonumber \\
&&-\frac{i\hbar }{2}\left[ Tr\;\ln G+Tr\;\ln F\right] +\Gamma _{Q}
\end{eqnarray}
where, if the position variable is explicit, $c_{11}=-c_{22}=1$, $%
c_{12}=c_{21}=0$,

\begin{equation}
D_{Aa,Bb}(u,v)=N\delta _{AB}\left[ c_{ab}\partial _{x}^{2}-c_{abc}\bar{\chi}%
^{c}\right] \delta (u,v),
\end{equation}
and $c_{abc}=1$ when all entries are $1$, $c_{abc}=-1$ when all entries are $%
2$, and $c_{abc}=0$ otherwise. When we use the compressed notation, it is
understood that $c_{ab}\equiv c_{ab}\delta (t_{a},t_{b})$ and $c_{abc}\equiv
c_{abc}\delta (t_{a},t_{b})\delta (t_{a},t_{c})$. $\Gamma _{Q}$ is the sum
of all $2PI$ vacuum bubbles with cubic vertices from $S_{3}$ and propagators
$G^{Aa,Bb}$ and $F^{ab}.$ Observe that $\Gamma _{Q}$ is independent of $\bar{%
\chi}^{c}$.

Taking variations of the $2PI$ $CTP$ $EA$ and identifying $\bar{\chi}^{1}=%
\bar{\chi}^{2}=\bar{\chi}$, we find the equations of motion

\begin{equation}
\frac{N}{2}\delta _{AB}D_{ab}-\frac{i\hbar }{2}\left[ G^{-1}\right] _{Aa,Bb}+%
\frac{1}{2}\Pi _{Aa,Bb}=0
\end{equation}

\begin{equation}
\frac{N}{2\lambda }c_{ab}-\frac{i\hbar }{2}\left[ F^{-1}\right] _{ab}+\frac{1%
}{2}\Pi _{ab}=0
\end{equation}

\begin{equation}
\frac{N}{\lambda }\left\{ \bar{\chi}\left( t\right) -M^{2}\right\} -\frac{N}{%
2}\delta _{AB}G^{A1,B1}(t,t)=0
\end{equation}
where $D_{ab}\left( u,v\right) =c_{ab}\left[ \partial _{x}^{2}-\bar{\chi}%
\left( u\right) \right] \delta (u,v)$,

\begin{equation}
\Pi _{Aa,Bb}=2\frac{\delta \Gamma _{Q}}{\delta G^{Aa,Bb}};\qquad \Pi _{ab}=2%
\frac{\delta \Gamma _{Q}}{\delta F^{ab}}
\end{equation}
We shall seek a solution with the structure

\begin{equation}
G^{Aa,Bb}=\frac{\hbar }{N}\delta ^{AB}G^{ab}(u,v)  \label{eq20}
\end{equation}
which is consistent with vanishing Noether charges. Then it is convenient to
write

\begin{equation}
F^{ab}=\frac{\hbar }{N}H^{ab};\qquad \Pi _{Aa,Bb}=\delta _{AB}P_{ab};\qquad
\Pi _{ab}\left( x,y\right) =NQ_{ab}\left( x,y\right)
\end{equation}
The equations become

\begin{equation}
D_{ab}-i\left[ G^{-1}\right] _{ab}+\frac{1}{N}P_{ab}=0  \label{schdy}
\end{equation}

\begin{equation}
\frac{1}{\lambda }c_{ab}-i\left[ H^{-1}\right] _{ab}+Q_{ab}=0  \label{eq23}
\end{equation}

\begin{equation}
\frac{1}{\lambda }\left\{ \bar{\chi}\left( t\right) -M^{2}\right\} -\frac{%
\hbar }{2}G^{11}(t,t)=0  \label{eq24}
\end{equation}
Observe that

\begin{equation}
P_{ab}=\frac{2}{\hbar }\frac{\delta \Gamma _{Q}}{\delta G^{ab}};\qquad
Q_{ab}=\frac{2}{\hbar }\frac{\delta \Gamma _{Q}}{\delta H^{ab}}
\end{equation}
These are the exact equations we must solve. The succesive $1/N$
approximations amount to different constitutive relations expressing $P_{ab}$
and $Q_{ab}$ in terms of the propagators.

The key observation is that in any given Feynman graph each vertex
contributes a power of $N,$ each internal line a power of
$N^{-1}$, and each trace over group indices another power of $N.$
We have both $G$ and $H$ internal lines, but the $G$ lines only
appear in closed loops. On each loop, the number of vertices
equals the number of $G$ lines, so there only remains one power
of $N$ from the single trace over group labels. Therefore the
overall power of the graph is the number of $G$ loops minus the
number of $H$ lines. Now, since we only consider $2PI$ graphs,
there is a minimum number
of $H$ lines for a given number of $G$ loops. For example, if there are two $%
G$ loops, they must be connected by no less than $3$ $H$ lines,
and so this graph cannot be higher than $NNLO.$ A graph with $3$
$G$ loops cannot have less than $5$ $H$ lines, and so on.

We conclude that $\Gamma _{Q}$ vanishes at $LO,$ and therefore $%
P_{ab}=Q_{ab}=0.$ There is only one $NLO$ graph, consisting of a single $G$
loop and a single $H$ line. This graph leads to

\begin{equation}
\Gamma _{Q}^{NLO}=\left( -i\hbar \right) \left( -\frac{1}{2}\right) \left( -%
\frac{N}{2Z_{0}\hbar }\right) ^{2}2N\left( \frac{\hbar }{N}\right)
^{3}c_{abc}c_{def}\int dudv\;H^{ad}\left( u,v\right) G^{be}\left( u,v\right)
G^{cf}\left( u,v\right)
\end{equation}
Therefore, we get

\begin{equation}
P_{ab}=i\hbar c_{acd}c_{bef}H^{ce}G^{df}  \label{eq75}
\end{equation}

\begin{equation}
Q_{ab}=\frac{i\hbar }{2}c_{acd}c_{bef}G^{ce}G^{df}  \label{eq76}
\end{equation}

\subsection{From correlations to the reduced density matrix}

The 2PIEA yields equations of motion for the (arbitrary time) two-point
functions of the theory. Given a solution of these equations, in principle
we may find the expectation values of a large family of composite operators
at any given time. Suppose we adopt a coarse-grained description where we
choose a certain number of these expectation values as the relevant
variables to describe the system. Then there will be a single density matrix
which has maximum von Neumann entropy with respect to the class of states
reproducing the preferred expectation values. This maximum entropy density
matrix is the reduced density matrix for the system, and its entropy its
correlation entropy. The $H$ theorem is the statement that the correlation
entropy grows in time, when the correlations themselves are evolved using
the equations derived from the $2PIEA$ truncated to some order in the $1/N$
expansion (this statement will be qualified below).

Let us consider first the simplest case where we describe the system by
specifying the values of $\left\langle x_{B}x_{B}\right\rangle ,$ $%
\left\langle x_{B}p_{B}+p_{B}x_{B}\right\rangle $ and $\left\langle
p^{B}p^{B}\right\rangle $ ($p^{A}=N\dot{x}^{A}$ is the momentum conjugate to
$x_{B}$) at every moment of time (by symmetry, we assume $\left\langle
x_{B}x_{C}\right\rangle =\delta _{BC}\left\langle x_{B}x_{B}\right\rangle /N$%
, and similarly in the other cases). This choice of relevant
variables does not utilize or display the full power of the $2PI$
$CTP$ $EA$, in the sense that the $2PI$ $%
CTP$ $EA$ yields equations of motion which, if carried beyond
$LO$, account for the build up of non-Gaussian correlations.
However, this restriction is assumed explicitly or implicitly in
most of the work on thermalization, and the focus is placed on
the shape of the spectrum of the two-point functions at different
times. The situation is analogous to Boltzmann theory, where the
equations of motion can describe the build-up of two particle
correlation functions from an uncorrelated initial state, but the
Boltzmann entropy is defined from the one particle distribution
function alone.

At every moment of time we define a maximum entropy density matrix

\begin{equation}
\rho \left( t\right) =Z^{-1}e^{-H_{0}}
\end{equation}
where

\begin{equation}
H_{0}={\frac{1}{2}} \left\{ \alpha p^{B}p^{B}+\beta \left(
x_{B}p_{B}+p_{B}x_{B}\right) +\gamma x_{B}x_{B}\right\}
\end{equation}
and

\begin{equation}
Z=\mathrm{Tr\ }e^{-H_{0}}=z^{N}
\end{equation}

\begin{equation}
z=\frac{1}{2\sinh \left[ \frac{1}{2}\hbar \sigma \right] }
\end{equation}
where

\begin{equation}
\sigma ^{2}=\alpha \gamma -\beta ^{2}
\end{equation}
(see Appendix A). The parameter $\sigma $ measures how far the system is
from a pure state. To see this, observe that

\begin{equation}
\mathrm{Tr}\rho ^{2}=\frac{\mathrm{Tr\;}e^{-2H_{0}}}{Z^{2}}=\left[ \frac{%
2\sinh ^{2}\left[ \frac{1}{2}\hbar \sigma \right] }{\sinh \left[ \hbar
\sigma \right] }\right] ^{N}=\left[ \tanh \left[ \frac{1}{2}\hbar \sigma
\right] \right] ^{N}
\end{equation}
So $\sigma \rightarrow \infty $ yields a pure state, while the state is
mixed for any finite $\sigma .$ Therefore, to show the $H$ theorem, we must
show that $d\sigma /dt<0.$

Let us begin by showing that $\sigma $ may be written directly in terms of
the expectation values for binary products of canonical variables. Indeed,
we have

\begin{equation}
-2\frac{\partial }{\partial \alpha }\ln Z=\left\langle
p^{B}p^{B}\right\rangle =\frac{N\hbar }{\tanh \left[ \frac{1}{2}\hbar \sigma
\right] }\frac{\gamma }{2\sigma }
\end{equation}

\begin{equation}
-2\frac{\partial }{\partial \gamma }\ln Z=\left\langle
x_{B}x_{B}\right\rangle =\frac{N\hbar }{\tanh \left[ \frac{1}{2}\hbar \sigma
\right] }\frac{\alpha }{2\sigma }
\end{equation}

\begin{equation}
-2\frac{\partial }{\partial \beta }\ln Z=\left\langle
x_{B}p_{B}+p_{B}x_{B}\right\rangle =\frac{-N\hbar }{\tanh \left[ \frac{1}{2}%
\hbar \sigma \right] }\frac{\beta }{\sigma }
\end{equation}
Therefore

\begin{equation}
4\left\langle p^{B}p^{B}\right\rangle \left\langle x_{B}x_{B}\right\rangle
-\left\langle x_{B}p_{B}+p_{B}x_{B}\right\rangle ^{2}=\left( \frac{N\hbar }{%
\tanh \left[ \frac{1}{2}\hbar \sigma \right] }\right) ^{2}
\end{equation}

As an aside, these formulae show that

\begin{equation}
\left\langle H_{0}\right\rangle =\frac{N\hbar \sigma }{2\tanh \left[ \frac{1%
}{2}\hbar \sigma \right] }
\end{equation}
therefore the von Neumann entropy is

\begin{equation}
S=-\left\langle \ln \rho \right\rangle =\left\langle H_{0}\right\rangle +\ln
Z=\frac{N\hbar \sigma }{2\tanh \left[ \frac{1}{2}\hbar \sigma \right] }-N\ln
\left[ \sinh \left[ \frac{1}{2}\hbar \sigma \right] \right] -N\ln 2
\end{equation}
Observe that

\begin{equation}
\frac{dS}{d\sigma }=-\frac{N\hbar ^{2}\sigma }{4\sinh ^{2}\left[ \frac{1}{2}%
\hbar \sigma \right] }<0  \label{desedete}
\end{equation}
So again, to obtain a $H$ theorem we must show that $\sigma $ is
non-increasing in time.

\subsection{The physical basis of the $H$ theorem}

It is clear that if we could solve the exact evolution, the $H$ theorem
would be manifest: Given $\rho \left( t_{0}\right) $ at the initial time $%
t_0 $, solve the exact Liouville equation up to a time $t_{1}$. Let $\bar{%
\rho}\left( t_{1}\right) $ be the result. We extract the new expectation
values from $\bar{\rho}\left( t_{1}\right) $ and use them to construct the
new maximum entropy density matrix $\rho \left( t_{1}\right) $. Then $%
S\left[ \rho \left( t_{1}\right) \right] \geq S\left[
\bar{\rho}\left( t_{1}\right) \right] ,$ by definition, and
$S\left[ \bar{\rho}\left( t_{1}\right) \right] =S\left[ \rho
\left( t_{0}\right) \right] $, because the exact evolution is
unitary, thus the $H$ theorem.

What we need to determine is whether, given the approximate
dynamics for the expectation value provided by the $1/N$ scheme,
the $H$ theorem still obtain. In fact, we know to $LO$ it does
not. From the angle of addressing the thermalization issue, the
$LO$ approximation totally misses the point.

Physically, we expect to obtain a $H$ theorem because our
description of the system is incomplete, in that it ignores
higher correlations. We may rightfully call this entropy
\textit{correlation entropy}, as explained in the Introduction.
The system as described by a finite order $1/N$ approximation is
an open system, in the following sense: Suppose we try to
reproduce it in the lab. In order to have the different
correlations evolving according to the $2PI$ $1/N$ equations
appropriate to the desired order, rather than the full Schwinger
- Dyson hierarchy, we must keep nudging it to conform to this
artificially created condition (again, due to our inability to
comprehend the complete picture). So there is energy and (in
principle) entropy flow in and out of the system besides
correlation entropy production. The sum of the entropy produced
in the system, and that exchanged with the environment (which
keeps the system on course) need not be positive.

In the $LO$ case, as the approximate evolution (described by a
quantum Vlasov equation \cite{KME}) is unitary, there is no net
entropy production and no $H$-theorem. This implies that the two
sources of entropy change must cancel exactly. Note that in such
a case, if one works with a Fock representation, for boson
fields, the number of particles increases with time and can be
used as a measure of field entropy (what Hu and Pavon \cite{HuPav}
called an intrinsic measure of field entropy, as described in Sec.
IIA above. See also \cite{HKMP96}.) However, this should not be
confused with the
correlation entropy of the Boltzmann kind under study for which the $H$%
-theorem is defined.

Of course, we expect better approximations to be closer to the
exact dynamics, therefore requiring less external control of (or
rather, tampering with) the system. This reduces the entropy loss
to the environment. Eventually correlation entropy production
becomes the dominant factor, and a $H$ theorem is obtained.

To summarize, the relevant question is not whether there is a $H$ theorem
for a given choice of relevant variables, which is obvious, but rather if
the $NLO$ approximation is good enough to make it manifest, or we must go
even higher.

\subsection{Statement of the $H$ theorem}

To continue our investigation, first note that

\begin{equation}
\Delta =\frac{d}{dt}\left\{ 4\left\langle p^{B}p^{B}\right\rangle
\left\langle x_{B}x_{B}\right\rangle -\left\langle
x_{B}p_{B}+p_{B}x_{B}\right\rangle ^{2}\right\} =-\left( \frac{N^{2}\hbar
^{2}}{\tanh \left[ \frac{1}{2}\hbar \sigma \right] }\right) \frac{1}{\sinh
^{2}\left[ \frac{1}{2}\hbar \sigma \right] }\frac{d\sigma }{dt}
\label{desigmadete}
\end{equation}
So we must show that the left hand side is positive. Since the
$1/N$ scheme proceeds via taking the expectation values of the
canonical equations of motion, we may use them to simplify this
expression. The canonical equations are (for simplicity, we use
the form without the auxiliary field)

\begin{equation}
\dot{x}^{A}=\frac{p^{A}}{N}
\end{equation}

\begin{equation}
\dot{p}^{A}=-N\left[ M^{2}+\frac{\lambda }{2}\left( x_{B}x_{B}\right)
\right] x^{A}
\end{equation}
Therefore

\begin{equation}
\frac{d}{dt}\left\langle x_{B}x_{B}\right\rangle =\frac{1}{N}\left\langle
x_{B}p_{B}+p_{B}x_{B}\right\rangle
\end{equation}

\begin{equation}
\frac{d}{dt}\left\langle p^{B}p^{B}\right\rangle =-N\left[ M^{2}\left\langle
x_{B}p_{B}+p_{B}x_{B}\right\rangle -\frac{\lambda }{2}\left\langle \left(
x_{C}x_{C}\right) x_{B}p_{B}+p_{B}x_{B}\left( x_{C}x_{C}\right)
\right\rangle \right]
\end{equation}
Recall the canonical commutation relations

\begin{equation}
x_{B}p_{B}=\frac{1}{2}\left( x_{B}p_{B}+p_{B}x_{B}\right) +\frac{i\hbar }{2}
\end{equation}

\begin{equation}
p_{B}x_{B}=\frac{1}{2}\left( x_{B}p_{B}+p_{B}x_{B}\right) -\frac{i\hbar }{2},
\end{equation}
so

\begin{eqnarray}
\frac{d}{dt}\left\langle p^{B}p^{B}\right\rangle &=&-N\left[
M^{2}\left\langle x_{B}p_{B}+p_{B}x_{B}\right\rangle \right.  \nonumber \\
&&\left. +\frac{\lambda }{4}\left\langle \left( x_{C}x_{C}\right) \left(
x_{B}p_{B}+p_{B}x_{B}\right) +\left( x_{B}p_{B}+p_{B}x_{B}\right) \left(
x_{C}x_{C}\right) \right\rangle \right] .
\end{eqnarray}
Finally

\begin{equation}
\frac{d}{dt}\left\langle x_{B}p_{B}+p_{B}x_{B}\right\rangle =\frac{2}{N}%
\left\langle p^{B}p^{B}\right\rangle -2N\left[ M^{2}\left\langle
x_{B}x_{B}\right\rangle +\frac{\lambda }{2}\left\langle \left(
x_{C}x_{C}\right) ^{2}\right\rangle \right] ,
\end{equation}
so

\begin{eqnarray}
\Delta &=&\frac{d}{dt}\left\{ 4\left\langle p^{B}p^{B}\right\rangle
\left\langle x_{B}x_{B}\right\rangle -\left\langle
x_{B}p_{B}+p_{B}x_{B}\right\rangle ^{2}\right\}  \nonumber \\
&=&\lambda N\left\{ 2\left\langle x_{B}p_{B}+p_{B}x_{B}\right\rangle
\left\langle \left( x_{C}x_{C}\right) ^{2}\right\rangle \right. \\
&&\left. -\left\langle x_{B}x_{B}\right\rangle \left\langle \left(
x_{C}x_{C}\right) \left( x_{B}p_{B}+p_{B}x_{B}\right) +\left(
x_{B}p_{B}+p_{B}x_{B}\right) \left( x_{C}x_{C}\right) \right\rangle \right\}
\end{eqnarray}

This may be rewritten in the following suggestive way

\begin{eqnarray}
\Delta &=&\lambda \left\{ 2\left\langle \left( x_{C}x_{C}\right)
^{2}\right\rangle \frac{d}{dt}\left\langle x_{B}x_{B}\right\rangle
-\left\langle x_{B}x_{B}\right\rangle \frac{d}{dt}\left\langle \left(
x_{C}x_{C}\right) ^{2}\right\rangle \right\}  \nonumber \\
&=&\left( -\lambda \right) \left\langle x_{B}x_{B}\right\rangle ^{3}\frac{d}{%
dt}\left[ \frac{\left\langle \left( x_{C}x_{C}\right) ^{2}\right\rangle }{%
\left\langle x_{B}x_{B}\right\rangle ^{2}}\right] ,  \label{delta}
\end{eqnarray}
so the $H$ theorem reads

\begin{equation}
\frac d{dt}\left[ \frac{\left\langle \left( x_Cx_C\right) ^2\right\rangle }{%
\left\langle x_Bx_B\right\rangle ^2}\right] \leq 0.
\end{equation}
Or, equivalently,

\begin{equation}
\frac{d}{dt}\left[ \frac{\left\langle \left( x_{C}x_{C}\right)
^{2}\right\rangle -\left\langle x_{B}x_{B}\right\rangle ^{2}}{\left\langle
x_{B}x_{B}\right\rangle ^{2}}\right] \leq 0  \label{teorema}
\end{equation}
Observe that this expression is invariant under a rescaling of the field.

\section{Correlation entropy production in weakly coupled theories}

The results from the last section may be summarized as (cfr. Eqs.
(\ref {desedete}), (\ref{desigmadete}) and (\ref{delta}))

\begin{equation}
\frac{dS}{dt}=\left( \frac{-\lambda }{4N}\right) \sigma \tanh \left[ \frac{1%
}{2}\hbar \sigma \right] \left\langle x_{B}x_{B}\right\rangle ^{3}\frac{d}{dt%
}\left[ \frac{\left\langle \left( x_{C}x_{C}\right) ^{2}\right\rangle }{%
\left\langle x_{B}x_{B}\right\rangle ^{2}}\right]   \label{resumen}
\end{equation}
We shall not attempt to give a general proof that this quantity is
non-negative, but only check that it is so in some simple cases.

We must first express the expectation value in Eq. (\ref{resumen})
to the required order in $1/N$. Recall that the $q-$number
auxiliary field $\chi $ was introduced as a formal Gaussian
process with correlation functions

\begin{equation}
\left\langle \chi \left( t\right) \right\rangle _{\chi }=M^{2}+\frac{\lambda
}{2}x_{B}x_{B}\left( t\right)
\end{equation}

\begin{equation}
\left\langle \chi \left( t\right) \chi \left( t^{\prime }\right)
\right\rangle _{\chi }=\left( M^{2}+\frac{\lambda }{2}x_{B}x_{B}\left(
t\right) \right) \left( M^{2}+\frac{\lambda }{2}x_{B}x_{B}\left( t^{\prime
}\right) \right) +\frac{i\lambda \hbar }{N}\delta \left( t-t^{\prime }\right)
\end{equation}
where $\left\langle {}\right\rangle _{\chi }$ denotes an
expectation value with respect to the $\chi $ variable alone.
Taking a further expectation value over the quantum state of the
$x$ fields, we recover Eq. (\ref{eq24}) and

\begin{equation}
\frac{\hbar }{N}H^{11}\left( t,t^{\prime }\right) =\frac{\lambda ^{2}}{4}%
\left[ \left\langle x_{B}x_{B}\left( t\right) x_{B}x_{B}\left( t^{\prime
}\right) \right\rangle -\left\langle x_{B}x_{B}\left( t\right) \right\rangle
\left\langle x_{B}x_{B}\left( t^{\prime }\right) \right\rangle \right] +%
\frac{i\lambda \hbar }{N}\delta \left( t-t^{\prime }\right)
\end{equation}
Therefore the $H$ theorem boils down to showing that

\begin{equation}
\frac{4\hbar }{\lambda ^{2}N}\frac{d}{dt}\frac{\left[ H^{11}\left(
t,t\right) -i\lambda \delta \left( 0\right) \right] }{\left\langle
x_{B}x_{B}\left( t\right) \right\rangle ^{2}}\leq 0
\end{equation}

It is clear that there is no correlation entropy production in free
theories, so if the $H$ theorem is to hold as a categorical relationship,
then it must hold already to lowest order in the coupling constant.
Therefore we may expand $H^{11}\left( t,t\right) $ in powers of $\lambda ,$
keeping only the lowest nontrivial order. Iterating the equation of motion

\begin{equation}
H^{ac}\left( t,t^{\prime }\right) =i\lambda c^{ac}\delta \left( t-t^{\prime
}\right) -\lambda \int ds\;c^{ab}Q_{bd}\left( t,s\right) H^{dc}\left(
s,t^{\prime }\right)
\end{equation}
we get

\begin{equation}
H^{11}\left( t,t\right) =i\lambda \delta \left( 0\right) -i\lambda
^{2}Q^{11}\left( t,t\right) +i\frac{\lambda ^{3}\hbar ^{2}}{4}\Psi
\end{equation}
where

\begin{eqnarray}
\Psi &=&\frac{4}{\hbar ^{2}}\int dt^{\prime }\;\left[ \left( Q_{11}\left(
t,t^{\prime }\right) \right) ^{2}-\left( Q_{12}\left( t,t^{\prime }\right)
\right) ^{2}\right]  \nonumber \\
&=&\int dt^{\prime }\;\left[ \left( G^{12}\left( t,t^{\prime }\right)
\right) ^{4}-\left( G^{11}\left( t,t^{\prime }\right) \right) ^{4}\right]
\end{eqnarray}

We shall investigate the $H$ theorem in two limiting cases, first at early
times for an arbitrary (diagonal in occupation number) initial state, and
then at late time for vacuum initial conditions.

\subsection{The $H$ theorem at early times}

To investigate the meaning of this expression, let us expand the $LO$
propagators in terms of mode functions

\begin{equation}
f\left( t\right) =\frac{1}{\sqrt{2\omega }}e^{-i\omega t}.
\end{equation}
Concretely, if the initial state is diagonal in occupation number, we have

\begin{equation}
G^{11}\left( t,t^{\prime }\right) =\frac{1}{2\omega }e^{-i\omega \left(
t_{>}-t_{<}\right) }+\frac{n}{\omega }\cos \omega \left( t-t^{\prime }\right)
\end{equation}

\begin{equation}
G^{12}\left( t,t^{\prime }\right) =\frac{1}{2\omega }e^{i\omega \left(
t-t^{\prime }\right) }+\frac{n}{\omega }\cos \omega \left( t-t^{\prime
}\right) ,
\end{equation}
where $n=\left\langle a^{\dagger }a\right\rangle $. Therefore

\begin{equation}
\Psi \left( t\right) =\frac{i}{8\omega ^{4}}\int_{0}^{t}dt^{\prime }\;\psi
\left( t-t^{\prime }\right)
\end{equation}

\begin{equation}
\psi \left( t\right) =4\left[ 2n^{3}+3n^{2}-n\right] \sin 2\omega t+\left[
2\left( 2n^{3}+3n^{2}\right) +4n+1\right] \sin 4\omega t
\end{equation}
(See Appendix B). Since

\begin{equation}
\left\langle x_{B}x_{B}\right\rangle =\frac{1+2n}{2\omega }
\end{equation}
is constant, the relevant derivative is $\left( i\right) d\Psi /dt,$ which
is proportional to $d\left( -\int \psi dt\right) /dt$. The $H$ theorem [$%
\left( i\right) d\Psi /dt<0$] will obtain if $\left( d/dt\right)
\int \psi dt>0$. Now, since $\psi \left( 0\right) =0$

\begin{equation}
\frac{d}{dt}\int_{0}^{t}dt^{\prime }\;\psi \left( t-t^{\prime }\right)
=\int_{0}^{t}dt^{\prime }\;\frac{d}{dt}\psi \left( t-t^{\prime }\right)
=-\int_{0}^{t}dt^{\prime }\;\frac{d}{dt^{\prime }}\psi \left( t-t^{\prime
}\right) =\psi \left( t\right)
\end{equation}
which is indeed positive at short times.

\subsection{The $H$ theorem at late times}

Let us assume vacuum initial conditions. The lowest order propagators may be
written in terms of mode functions

\begin{equation}
G^{11}\left( t,t^{\prime }\right) =F\left( t_{>}\right) F^{*}\left(
t_{<}\right)
\end{equation}

\begin{equation}
G^{12}\left( t,t^{\prime }\right) =F^{*}\left( t\right) F\left( t^{\prime
}\right) .
\end{equation}
\begin{equation}
\Psi =\int_{0}^{t}dt^{\prime }\;\left\{ F^{4}\left( t^{\prime }\right)
F^{*4}\left( t\right) -F^{4}\left( t\right) F^{*4}\left( t^{\prime }\right)
\right\} .
\end{equation}
We have

\begin{equation}
\left\langle x_{B}x_{B}\right\rangle =\hbar \left| F\left( t\right) \right|
^{2}
\end{equation}
So the relevant inequality is

\begin{equation}
i\frac{d}{dt}\frac{1}{\left| F\left( t\right) \right| ^{4}}%
\int_{0}^{t}dt^{\prime }\;\left\{ F^{4}\left( t^{\prime }\right)
F^{*4}\left( t\right) -F^{4}\left( t\right) F^{*4}\left( t^{\prime }\right)
\right\} \leq 0
\end{equation}
Or, equivalently,

\begin{equation}
J=i\frac{d}{dt}\int_{0}^{t}dt^{\prime }\;\left\{ \frac{F^{4}\left( t^{\prime
}\right) }{F^{4}\left( t\right) }-\frac{F^{\ast 4}\left( t^{\prime }\right)
}{F^{\ast 4}\left( t\right) }\right\} \leq 0.
\end{equation}
Now

\begin{equation}
J=-4i\int_{0}^{t}dt^{\prime }\;\left\{ F^{^{\prime }}\left( t\right) \frac{%
F^{4}\left( t^{\prime }\right) }{F^{5}\left( t\right) }-F^{*^{\prime
}}\left( t\right) \frac{F^{*4}\left( t^{\prime }\right) }{F^{*5}\left(
t\right) }\right\}
\end{equation}
The mode functions may be expanded in terms of WKB modes

\begin{equation}
F\left( t\right) =\alpha \left( t\right) f\left( t\right) +\beta \left(
t\right) f^{*}\left( t\right)
\end{equation}

\begin{equation}
F^{\prime }\left( t\right) =\alpha \left( t\right) f^{\prime }\left(
t\right) +\beta \left( t\right) f^{*^{\prime }}\left( t\right)
\end{equation}

At long times, the integrals will be dominated by the non-oscillatory terms

\begin{equation}
F^{4}\left( t^{\prime }\right) \sim 6\alpha ^{2}\left( t\right) \beta
^{2}\left( t\right) \left| f\left( t\right) \right| ^{4}
\end{equation}

\begin{equation}
J\sim \frac{-24it}{\left| F\left( t\right) \right| ^{10}}\;\left| f\left(
t\right) \right| ^{4}\left\{ F^{^{\prime }}\left( t\right) F^{*5}\left(
t\right) \alpha ^{2}\left( t\right) \beta ^{2}\left( t\right) -F^{*^{\prime
}}\left( t\right) F^{5}\left( t\right) \alpha ^{*2}\left( t\right) \beta
^{*2}\left( t\right) \right\}.
\end{equation}
We make the analogous approximation

\begin{eqnarray}
F^{^{\prime }}\left( t\right) F^{*5}\left( t\right) &=&\alpha \left(
t\right) f^{\prime }\left( t\right) F^{*5}\left( t\right) +\beta \left(
t\right) f^{*^{\prime }}\left( t\right) F^{*5}\left( t\right)  \nonumber \\
&\sim &10\left| f\left( t\right) \right| ^{4}\alpha ^{*2}\left( t\right)
\beta ^{*2}\left( t\right) \left\{ \left| \alpha \left( t\right) \right|
^{2}f^{\prime }\left( t\right) f^{*}\left( t\right) +\left| \beta \left(
t\right) \right| ^{2}f^{*^{\prime }}\left( t\right) f\left( t\right) \right\}
\end{eqnarray}
and get

\begin{equation}
J\sim \frac{-240it}{\left| F\left( t\right) \right| ^{10}}\left| f\left(
t\right) \right| ^{8}\left| \alpha \left( t\right) \right| ^{4}\left| \beta
\left( t\right) \right| ^{4}\left\{ \left| \alpha \left( t\right) \right|
^{2}-\left| \beta \left( t\right) \right| ^{2}\right\} \left\{ f^{\prime
}\left( t\right) f^{*}\left( t\right) -f^{*^{\prime }}\left( t\right)
f\left( t\right) \right\}.
\end{equation}
But

\begin{equation}
\left| \alpha \left( t\right) \right| ^{2}-\left| \beta \left( t\right)
\right| ^{2}=1
\end{equation}

\begin{equation}
f^{\prime }\left( t\right) f^{*}\left( t\right) -f^{*^{\prime }}\left(
t\right) f\left( t\right) =-i.
\end{equation}
So

\begin{equation}
J\sim \frac{-240t}{\left| F\left( t\right) \right| ^{10}}\left| f\left(
t\right) \right| ^{8}\left| \alpha \left( t\right) \right| ^{4}\left| \beta
\left( t\right) \right| ^{4}<0
\end{equation}

$\mathcal{QED}$.

The existence of a H-theorem at the NLO is as reassuring as the
absence thereof at the LO. It is also interesting to observe that
entropy production is (formally) a quantity of the order of
$N^{-2}$; a strict (formal) $1/N$ expansion would neglect such a
quantity, but it would be generated in a numerical solution of
the full $1/N$ equations of motion. Therefore, it is suggestive
that the statistical mechanical description of an interacting
quantum field at the NLO level of approximation to a 2PI CTP
effective action may permit ``thermalization'', as claimed by
\cite {BC01,B02} based on numerical results. Note that our
analysis corroborating this claim is predicated upon the
restrictive conditions discussed above. Extension of this work to
full field theory in conjuction with the reported results of
\cite{CH02} is under investigation.\\

\noindent \textbf{Acknowledgements} We thank Steve Ramsey for
collaboration at the beginning stage of this work, some
preliminary results were recorded in Chapter 5 of his University
of Maryland Ph. D. thesis \cite{RamseyPhD}. We acknowledge
discussions of NLO scheme with Emil Mottola, Paul Anderson and
Fred Cooper. EC is supported in part by CONICET, UBA,
Fundaci\'{o}n Antorchas and ANPCyT. BLH is supported in part by
NSF grant PHY98-00967.

\section{Appendix A: Computing the partition function}

To compute $Z,$ it is best to diagonalize $H_{0}.$ Introduce new canonical
variables

\begin{equation}
\xi _A=\left( \cosh u\right) p_A+\left( \sinh u\right) x_A
\end{equation}

\begin{equation}
\eta _A=\left( \sinh u\right) p_A+\left( \cosh u\right) x_A
\end{equation}

so that $[\xi _A,\eta _B] =[ p_A,x_B ] =( -i\hbar) \delta _{AB}.$ Then

\begin{equation}
p_A=\left( \cosh u\right) \xi _A-\left( \sinh u\right) \eta _A
\end{equation}

\begin{equation}
x_A=-\left( \sinh u\right) \xi _A+\left( \cosh u\right) \eta _A
\end{equation}
and

\begin{equation}
H_0=({\frac 12})\left\{ A\xi ^B\xi ^B+B\left( \eta _B\xi _B+\xi _B\eta
_B\right) +C\eta _B\eta _B\right\}
\end{equation}
where

\begin{equation}
A=\alpha \cosh ^2u+\gamma \sinh ^2u-2\beta \cosh u\sinh u
\end{equation}

\begin{equation}
B=-\left( \gamma +\alpha \right) \cosh u\sinh u+\beta \left( \cosh
^{2}u+\sinh ^{2}u\right)
\end{equation}

\begin{equation}
C=\alpha \sinh ^{2}u+\gamma \cosh ^{2}u-2\beta \cosh u\sinh u
\end{equation}

To diagonalize the Hamiltonian, we require $B=0,$ namely

\begin{equation}
\tanh 2u=\frac{2\beta }{\gamma +\alpha }
\end{equation}
Now $Z$ is the usual partition function for $N$ harmonic oscillators with
frequency $\omega ^{2}=C/A$ at inverse temperature $A$.

\section{Appendix B: $\psi \left( t\right) $}

\begin{eqnarray}
\Psi \left( t\right) &=&\frac{1}{16\omega ^{4}}\int_{0}^{t}dt^{\prime
}\;\left\{ \left[ e^{i\omega \left( t-t^{\prime }\right) }+2n\cos \omega
\left( t-t^{\prime }\right) \right] ^{4}-\left[ e^{-i\omega \left(
t-t^{\prime }\right) }+2n\cos \omega \left( t-t^{\prime }\right) \right]
^{4}\right\}  \nonumber \\
\ &=&\frac{i}{8\omega ^{4}}\int_{0}^{t}dt^{\prime }\;\left\{ 32n^{3}\cos
^{3}\omega \left( t-t^{\prime }\right) \sin \omega \left( t-t^{\prime
}\right) +24n^{2}\cos ^{2}\omega \left( t-t^{\prime }\right) \sin 2\omega
\left( t-t^{\prime }\right) \right.  \nonumber \\
&&\ \ \left. +8n\cos \omega \left( t-t^{\prime }\right) \sin 3\omega \left(
t-t^{\prime }\right) +\sin 4\omega \left( t-t^{\prime }\right) \right\}
\nonumber \\
\ &=&\frac{i}{8\omega ^{4}}\int_{0}^{t}dt^{\prime }\;\left\{ 8\left(
2n^{3}+3n^{2}\right) \cos ^{2}\omega \left( t-t^{\prime }\right) \sin
2\omega \left( t-t^{\prime }\right) \right.  \nonumber \\
&&\ \ \left. +8n\cos \omega \left( t-t^{\prime }\right) \sin 3\omega \left(
t-t^{\prime }\right) +\sin 4\omega \left( t-t^{\prime }\right) \right\}
\nonumber \\
\ &=&\frac{i}{8\omega ^{4}}\int_{0}^{t}dt^{\prime }\;\left\{ 2\left(
2n^{3}+3n^{2}\right) \left[ 2\sin 2\omega \left( t-t^{\prime }\right) +\sin
4\omega \left( t-t^{\prime }\right) \right] \right.  \nonumber \\
&&\ \ \left. +4n\left[ \sin 4\omega \left( t-t^{\prime }\right) -\sin
2\omega \left( t-t^{\prime }\right) \right] +\sin 4\omega \left( t-t^{\prime
}\right) \right\}  \nonumber \\
\ &=&\frac{i}{8\omega ^{4}}\int_{0}^{t}dt^{\prime }\;\left\{ 4\sin 2\omega
\left( t-t^{\prime }\right) \left[ 2n^{3}+3n^{2}-n\right] +\sin 4\omega
\left( t-t^{\prime }\right) \left[ 2\left( 2n^{3}+3n^{2}\right) +4n+1\right]
\right\}
\end{eqnarray}\\

\end{document}